\documentclass[conference]{IEEEtran}
\usepackage{balance}
\usepackage[a4paper, left=1.27cm, right=1.27cm, top=1.815cm, bottom=4.3cm]{geometry}

\usepackage{float}
\usepackage{standard}
\usepackage{etex}

\usepackage{rotate}
\usepackage{algorithmic}
\usepackage{algorithm}
\usepackage{float}
\usepackage{tikz}
\usepackage{refcount}

\usepackage{pgf,tikz}
\usepackage{pgfplots}
\usepackage{pgfplotstable}
\usepackage{bm}
\usepackage{adjustbox}
\usetikzlibrary{calc}
\usepackage{relsize}
\usepackage{ifthen}

\usetikzlibrary{calc,fit,arrows,plotmarks,intersections,patterns,shapes,decorations,positioning, backgrounds, matrix}

\usepackage[numbers,sort&compress]{natbib}
\usepackage[ngerman, english]{babel}

\usetikzlibrary{arrows.meta}
\edef\wdArrowLength{2}
\tikzset{>={Latex[width=1.5mm,length=\wdArrowLength mm]}}
\usepackage{graphicx}

\usepackage{cancel}
\usepackage{mathtools}
\usepackage{shadethm}
\usepackage{empheq}
\usepackage{skull}
\usepgfplotslibrary{units}
\usepackage{calc}
\usepackage{nicefrac}
\usepackage{fancybox}
\usepackage{psfrag}
\usepackage{dsfont}

\title{When Every Symbol Counts: Resilient Wireless Systems Under Finite Blocklength Constraints}
\author{\IEEEauthorblockN{Kevin Weinberger$^{}$, and Aydin Sezgin$^{}$}
\IEEEauthorblockA{Institute of Digital Communication Systems, Ruhr University Bochum, Germany\\
Email: {\{kevin.weinberger,aydin.sezgin\}@rub.de} }
\thanks{This work was supported in part by the German Research Foundation (DFG) in the course of the project SPP2433 under the project no. 541021107 (Measurement Technology on Flying Platforms) under grant SE 1697/22-1.}
}
\date{\today}

\usepackage{graphicx}
\usepackage{placeins}
\usepackage{booktabs}
\usepackage{marvosym}
\usepackage{multirow}

\usepackage{latexsym, amsmath, amssymb, amsfonts, upgreek}
\usepackage[nolist]{acronym}

\usepackage{tikz}
\usetikzlibrary{calc,arrows,positioning,decorations,shadows,shapes,fadings,matrix}
\tikzset{>=latex'}
\tikzset{semithick}

\makeatletter
\providecommand{\IfElsePackageLoaded}[3]{\@ifpackageloaded{#1}{#2}{#3}}
\makeatother

\usepackage{subfigure}
\IfElsePackageLoaded{subfig}
	{\usepackage[subfigure]{tocloft}}{	
	\IfElsePackageLoaded{subfigure}
		{\usepackage[subfigure]{tocloft}}
		{\usepackage{tocloft}}
	}

\makeatletter

\def\tikz@delimiter#1#2#3#4#5#6#7#8{%
	\bgroup
		\pgfextra{\let\tikz@save@last@fig@name=\tikz@last@fig@name}%
		node[outer sep=0pt,inner sep=0pt,draw=none,fill=none,anchor=#1,at=(\tikz@last@fig@name.#2),#3]
		{%
			{\nullfont\pgf@process{\pgfpointdiff{\pgfpointanchor{\tikz@last@fig@name}{#4}}{\pgfpointanchor{\tikz@last@fig@name}{#5}}}}%
			\delimitershortfall\z@% as suggested by morbusg (maximum space not covered by a delimiter = 0)
			\resizebox*{!}{#8}{$\left#6\vcenter{\hrule height .5#8 depth .5#8 width0pt}\right#7$}%
		}
		\pgfextra{\global\let\tikz@last@fig@name=\tikz@save@last@fig@name}%
	\egroup%
}

\tikzset{hexagon/.code={
	\draw (0,2) -- (-4,0) -- (0,-2) -- (4,0) -- (0,2);
}}

\tikzset{phone/.code={
   \node [rectangle,rounded corners=1.5pt,draw,minimum height=0.6cm, minimum width=0.35cm] at (0,0){};
   \node [rectangle,rounded corners=1.5pt,draw,minimum height=0.5cm, minimum width=0.3cm] at (0,0){};
}}

\makeatletter
\def\cantox@vector#1#2#3#4#5#6#7#8{%
  \dimen@.5\p@
  \setbox\z@\vbox{\boxmaxdepth.5\p@
   \hbox{\kern-1.2\p@\kern#1\dimen@$#7{#8}\m@th$}}%
  \ifx\canto@fil\hidewidth  \wd\z@\z@ \else \kern-#6\unitlength \fi
  \ooalign{%
    \canto@fil$\m@th \CancelColor
    \vcenter{\hbox{\dimen@#6\unitlength \kern\dimen@
      \multiply\dimen@#4\divide\dimen@#3 \vrule\@depth\dimen@\@width\z@
      \vector(#3,-#4){#5}%
    }}_{\raise-#2\dimen@\copy\z@\kern-\scriptspace}$%
    \canto@fil \cr
    \hfil \box\@tempboxa \kern\wd\z@ \hfil \cr}}
\def\bcancelto#1#2{\let\canto@vector\cantox@vector\cancelto{#1}{#2}}
\makeatother

\IEEEoverridecommandlockouts
\usepackage{algorithmic}
\begin{document}
\maketitle
\begin{abstract}
As 6G evolves, wireless networks become essential for critical operations and enable innovative applications that demand seamless adaptation to dynamic environments and disruptions. Because these vital services require uninterrupted operation, their resilience to unforeseen disruptions is essential. However, implementing resilience necessitates rapid recovery procedures, which operate in the finite blocklength (FBL) regime, where short packets and added error-correction overhead can severely degrade communication efficiency. Due to this performance loss, always attempting recovery can backfire and result in worse outcomes than simply enduring the disruption under longer blocklengths.
In this work, we study these effects of FBL constraints within a resilience framework, incorporating reconfigurable intelligent surfaces (RIS) to enhance adaptation capabilities. By actively shaping the wireless environment, RIS help counteract some of the performance losses caused by FBL, enabling more effective recovery from disruptions. Numerical results reveal two critical blocklength thresholds: the first enables full recovery from the FBL penalty, while the second, at a higher blocklength, allows the system to recover from both the FBL penalty and the initial disruption, yielding a significant improvement in resilience performance. Additionally, we show that the number of RIS elements shifts these thresholds, enabling faster reconfiguration with shorter blocklengths and providing insights to the trade-offs between rate, blocklength, and reconfiguration effort under FBL conditions.
%As 6G and beyond redefine connectivity, wireless networks become the foundation of critical operations, making resilience more essential than ever. With this shift, wireless systems cannot only take on vital services previously handled by wired infrastructures but also enable novel innovative applications that would not be possible with wired systems. As a result, there is a pressing demand for strategies that can adapt to dynamic channel conditions, interference, and unforeseen disruptions, ensuring seamless and reliable performance in an increasingly complex environment. Despite considerable research, existing resilience assessments lack comprehensive key performance indicators (KPIs), especially those quantifying its adaptability, which are vital for identifying a system's capacity to rapidly adapt and reallocate resources. In this work, we bridge this gap by proposing a novel framework that explicitly quantifies the adaption performance by augmenting the gradient of the system's rate function. To further enhance the network resilience, we integrate Reconfigurable Intelligent Surfaces (RISs) into our framework due to their capability to dynamically reshape the propagation environment while providing alternative channel paths. Numerical results show that gradient augmentation enhances resilience by improving adaptability under adverse conditions while proactively preparing for future disruptions.
\end{abstract} 
%\begin{IEEEkeywords}
%
%%resilience, reconfigurable intelligent surface (ris), intelligent reflecting surface (irs), cloud radio access network (c-ran) resource allocation, quality of service.
%\end{IEEEkeywords}
\thispagestyle{empty}
\pagestyle{empty}

\section{Introduction}

As the vision of \ac{6G} networks takes shape, wireless infrastructure is expected to support not only human-centric applications but also a wide range of autonomous and AI-driven services requiring ultra-reliable and low-latency communication \cite{you2021towards}. Emerging use cases such as autonomous vehicles, industrial automation, and remote medical operations impose stringent \ac{QoS} requirements. In these scenarios, even brief communication disruptions can compromise safety, operational integrity, or user experience \cite{ChaccourAOI}.

To address these demands, next-generation wireless systems must go beyond traditional robustness and actively exhibit resilience, the capacity to absorb disruptions, adapt in real time, and recover functionality with minimal performance degradation \cite{brosInArms,RobertRes,ResByDesign}. This entails not only sustaining service during adverse events but also ensuring rapid and autonomous reconfiguration of resources within extremely short time spans \cite{accelRec,AntifragileRS}.

In such time-sensitive situations, the reorganization of the system's resources must involve \ac{FBL} transmissions, where short packets are used to meet latency constraints during the recovery efforts. Unlike the \ac{IBL} regime, \acp{FBL} introduce a trade-off between reliability, latency, and throughput due to limited channel uses and increased coding overhead \cite{FBL_Polyanskiy}. Consequently, adapting to disruptions under \ac{FBL} constraints can lead to efficiency losses, and in some cases, shortening the blocklengths during the recovery procedure may degrade system performance more than maintaining operation under degraded conditions. Thus, the \ac{FBL} regime is not merely a side effect of fast adaptation, it is a defining constraint that must be accounted for in the resilience design of future wireless systems.

To investigate this challenge, we develop a resilience framework that explicitly incorporates \ac{FBL} effects during the system's recovery phase. The utilized model quantifies resilience in terms of three core capabilities \cite{RobertRes}: absorption, adaptation, and time-to-recovery, allowing us to study the performance trade-offs between reacting to an outage or continuing operation with limited resources.

To mitigate the performance loss caused by \ac{FBL} constraints, the physical layer must be made more responsive to environmental changes. One promising technology in this context is the use of \ac{RIS} \cite{basar2019wireless, SynBenefits}, which allow for the dynamic control of the wireless propagation environment. By actively shaping the channel, RIS can establish alternative transmission paths or enhance existing links, improving resilience potential \cite{RIS_RES, CognitiveResilience} without requiring additional transmit power or bandwidth. This capability is especially valuable under \ac{FBL} conditions, where short transmission durations demand rapid and efficient adaptation \cite{RIS_FBL,RIS_FBLNurul}. In such time-constrained scenarios, the real-time reconfigurability of RIS becomes a critical enabler for maintaining service continuity, rerouting disrupted links, and sustaining communication quality during recovery.

Through this framework, we analyze when recovery should be initiated and when it is more beneficial to absorb disruptions without adapting. We demonstrate that the interplay between blocklength, RIS size and \ac{QoS} demands determines the feasibility and success of recovery. As it turns out, our framework reveals that there are critical blocklength thresholds, where the system becomes capable of not only compensating for FBL penalties but also recovering from disruptions.

\section{System Model}\label{ch:Sysmod}
\begin{figure}
	\centering \includegraphics[width=1\linewidth]{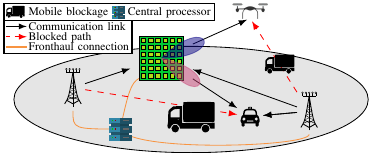}%2.75
	\caption{System model where mobile blockages may obstruct the direct AP-user links within a channel coherence interval, necessitating rapid adaptation of transmission strategies under finite blocklength constraints.}
	\label{fig:1}
\end{figure}%

%\textcolor{red}{This paper considers the \ac{RIS}-aided cell-free \ac{MIMO} downlink system depicted in Fig~\ref{fig:1}. In details, a set of single-antenna users $\mathcal{K}=\{1,\cdots,K\}$ is being served by a set of $L$-antenna \acp{AP} $\mathcal{N}=\{1,\cdots,N\}$. To enable an extra set of channel-links, we deploy an $M$-element \ac{RIS}, configured as a uniform planar array, within user range. The position is chosen in a strategic way, so that the \ac{RIS} is capable of providing alternative paths should direct links become blocked. The \ac{RIS} and the \acp{AP} are connected to and managed by a \ac{CP} via orthogonal fronthaul links. To meet the demands of the users, a \ac{QoS} target is set as a desired data rate denoted as $r_k^\mathsf{des}$, for each user.}
This work investigates a cell-free \ac{MIMO} downlink system enhanced by a \ac{RIS}, as illustrated in Fig.~\ref{fig:1}. Specifically, a group of single-antenna users $\mathcal{K} = \{1, \cdots, K\}$ is served by multiple \acp{AP}, each equipped with $L$ antennas, indexed by the set $\mathcal{N} = \{1, \cdots, N\}$. To introduce additional propagation paths, an $M$-element \ac{RIS}, arranged in a uniform planar array, is deployed within the center of the served area. Its placement is carefully selected to ensure that it can establish alternative links in the event that the direct AP-user channels become obstructed. The \ac{RIS} and all \acp{AP} are connected to a central processing unit (\ac{CP}) via orthogonal fronthaul connections. To fulfill user requirements, each user $k$ is assigned a quality-of-service (\ac{QoS}) target in the form of a desired data rate, denoted by $r_k^\mathsf{des}$.

\subsection{Channel Model}\label{ssec:chan}
\newcommand{\RisChanFull}[1]{\vect{h}_{#1} + \mat{G}_{#1}\vect{v}}
\newcommand{\RisChan}[1]{\vect{h}^\mathsf{eff}_{#1}}
The channel model adopts a quasi-static block fading assumption, where the channel coefficients remain constant over a coherence interval of duration $T_\mathsf{c}$ and vary independently between different coherence blocks. The transmit signal vector from the $n$-th \ac{AP} is defined as
\[
\vect{x}_{n} = \sum_{k \in \mathcal{K}} \vect{w}_{n,k} s_k, \quad \forall n \in \mathcal{N},
\]
where $s_k$ denotes the symbol intended for user $k$, drawn from an i.i.d. Gaussian codebook, and $\vect{w}_{n,k} \in \mathbb{C}^{L \times 1}$ is the beamforming vector provided by the central processor (\ac{CP}). The per-AP transmit signal is constrained by a maximum power limit:
\begin{align}\label{eq:powConst}
\mathbb{E}\{\vect{x}_n^H \vect{x}_n\} = \sum_{k \in \mathcal{K}} \| \vect{w}_{n,k} \|_2^2 \leq P^{\mathsf{max}}_n, \quad \forall n \in \mathcal{N}.
\end{align}

The aggregate transmit signal vector across all \acp{AP} can then be expressed as $
\vect{x} = [\vect{x}_1^T, \vect{x}_2^T, \dots, \vect{x}_N^T]^T \in \mathbb{C}^{NL \times 1}.
$
The direct channel between AP $n$ and user $k$ is denoted as $\vect{h}_{n,k} \in \mathbb{C}^{L \times 1}$. For RIS-assisted communication, the reflected channel is modeled as
\[
\mat{G}_{n,k} = \mat{H}_{n} \mathsf{diag}(\vect{v}) \vect{g}_k \in \mathbb{C}^{L \times 1},
\]
where $\mat{H}_n \in \mathbb{C}^{L \times M}$ is the AP-to-RIS channel, $\vect{g}_k \in \mathbb{C}^{M \times 1}$ is the RIS-to-user channel, and $\mathsf{diag}(\vect{v})$ contains the RIS reflection coefficients $\vect{v} = [v_1, v_2, \dots, v_M]^T$, with each $v_m = e^{j\theta_m}$ and $\theta_m \in [0, 2\pi]$. We define the aggregated direct channel vector for user $k$ as
$
\vect{h}_k = [\vect{h}_{1,k}^T, \vect{h}_{2,k}^T, \dots, \vect{h}_{N,k}^T]^T \in \mathbb{C}^{NL \times 1},
$
and the full AP-to-RIS channel matrix as
$
\vect{H} = [\mat{H}_1^T, \mat{H}_2^T, \dots, \mat{H}_N^T]^T \in \mathbb{C}^{NL \times M}.
$

Using these definitions, the effective received signal at user $k$ is given by the superposition of direct and RIS-reflected paths:
\begin{align}\label{recSgn}
y_k = (\vect{h}_k + \mat{G}_k \vect{v})^H \vect{x} + n_k = (\vect{h}_k^\mathsf{eff})^H \vect{x} + n_k,
\end{align}
where $\mat{G}_k = \vect{H} \mathsf{diag}(\vect{g}_k)$, $\vect{h}_k^\mathsf{eff} = \vect{h}_k + \mat{G}_k \vect{v}$, and $n_k \sim \mathcal{CN}(0, \sigma_k)$ is complex \ac{AWGN}.

\noindent
The received signal (\ref{recSgn}) at user $k$ is then given by
\begin{align}
y_k = (\RisChan{k})^H \vect{w}_k s_k + \sum_{i\in\mathcal{K}\setminus\{k\}} (\RisChan{k}) \vect{w}_i s_i + n_k,
\end{align}
where the first term represents user $k$'s desired signal, and the second term accounts for all the other user's interference.

Thus, we formulate the \ac{SINR} of user $k$ decoding its message as
\begin{align}
\Gamma_k = \frac{|(\RisChan{k})^H \vect{w}_k|^2}{\sum_{i\in\mathcal{K}\setminus\{k\}}|(\RisChan{k})^H \vect{w}_i|^2 + \sigma^2},
\end{align}
where $\sigma^2$ denotes the noise power.

Using these definitions, each user's \ac{QoS} demands is satisfied, if the following condition holds:
\begin{align}
R_k^\mathsf{des} \leq R_k \leq B \log_2(1+\Gamma_k) ,
\end{align}
where $B$ denotes the transmission bandwidth and $R_k$ represents the allocated rate of user $k$ under \ac{IBL} assumption. In case of a disruption, however, the system needs to reallocate resources quickly and consequently operates with \acp{FBL} of size $\eta$. According to \cite{FBL_Polyanskiy}, the QoS demands in the \ac{FBL}-regime are satisfied, if the following condition holds
\begin{align}\label{FBL_rate}
r_k^\mathsf{des} \leq  r_k \leq B \big(\log_2(1+\Gamma_k) -  \Omega \, \sqrt{{V(\Gamma_k)}/{\eta}} \big),
\end{align}
where $\Omega = Q^{-1}(\epsilon) \log_2(e)$, $\eta$ is the blocklength, $\epsilon$ is the \ac{BLER} and the function $Q^{-1}(\cdot)$ represents the inverse of the Gaussian $\mathcal{Q}$ function\footnote{$Q(x) = \int_{x}^{\infty} \frac{1}{\sqrt{2\pi}}\exp{(\frac{t^2}{2})}dt.$}. Moreover, $V(\Gamma_k)$ is the channel dispersion parameter given by \begin{align}\label{FBL_disp}
  V(\Gamma_k) = 1 - (1+\Gamma_k)^{-2},
\end{align}
which is justified under the assumption of Gaussian signaling in our model.
\newcommand{\RisOnlyChan}[1]{\vect{h}_{#1}^\mathsf{RIS}}

%When the direct links to user $k$ are blocked, by denoting $\RisOnlyChan{k} = \mat{G}_k\vect{v}$, we can define the RIS-link \ac{SINR} $\Gamma_k^\mathsf{RIS}$ and rate expressions $r_k^\mathsf{RIS}$ as
%\begin{align}
%&\Gamma_k^\mathsf{RIS} = \frac{|(\RisOnlyChan{k})^H \vect{w}_k|^2}{\sum_{i\in\mathcal{K}\setminus\{k\}}|(\RisOnlyChan{k})^H \vect{w}_i|^2 + \sigma^2}, \\
%&r_k^\mathsf{RIS} \leq B \log_2(1+\Gamma_k^\mathsf{RIS}),\label{RIS}
%\end{align}
%which effectively represents the case, in which user $k$ is only served over the RIS-assisted links. Since we quantify the network's adaptability with the rate's gradient, by denoting $a_{k,i}=\vect{v}^H \mat{G}_k^H \vect{w}_i$ and $\vect{b}_{k,i} = \vect{w}_i^H \mat{G}_k\vect{v}\mat{G}_k^H\vect{w}_i\in \mathbb{C}^{M\times 1}$ we can express the gradient of the RIS-link rate \ac{w.r.t.} the phase shifts as
%\begin{align}\label{gradV}
%\nabla_\vect{v}r_k^\mathsf{RIS} =  \frac{2B(\vect{b}_{k,k} - \Gamma_k^\mathsf{RIS} \vect{b}_{k,i})}{\ln(2)(\sum_{i\in \mathcal{K}}{|a_{k,i}|^2} + \sigma^2)}.
%\end{align}
%\textit{Proof:} For a detailed derivation, we refer to Appendix \ref{sec:AppA} 
%\input{sections/transmission_scheme.tex}
\section{Resilience in the Finite Blocklength Regime}

\subsection{Finite Blocklength as a Resilience Constraint}
Resilient wireless networks are expected to remain operational even under unpredictable disruptions, such as link blockages or resource variability. Ideally, recovery mechanisms would activate immediately upon detecting a disruption. However, in practice, particularly during rapid reconfiguration or rerouting, communication must resume over very short timescales. These urgent recovery phases restrict the number of symbols that can be transmitted, forcing the system to operate in the \ac{FBL} regime. Unlike the idealized \ac{IBL} assumption, \ac{FBL} imposes a fundamental trade-off: short codewords require increased error protection, thereby reducing the effective data rate and complicating recovery. This constraint directly impacts resilience, as limited transmission efficiency under \ac{FBL} not only delays system restoration but also introduces an overall performance degradation.

This degradation raises a critical decision: whether the system should attempt to recover or ignore the disruption altogether. If the system opts to switch to shorter blocklengths to initiate recovery, the additional penalties from \ac{FBL} coding may result in less information being transmitted than before the disruption. Since physical layer resources are shared among all users, switching to \ac{FBL} affects every user simultaneously. It follows that if the available resources are insufficient to overcome the \ac{FBL}-induced performance loss, the system risks ending up in a worse state than if it had simply ignored the blockage. A resilient network must therefore be capable of intelligently deciding whether to engage in recovery actions or to temporarily tolerate the disruption, balancing the trade-offs between adaptation benefits and the potential costs under \ac{FBL} constraints.

\subsection{Resilience Metric}
To quantify the trade-offs that the recovery under \ac{FBL}-constraints imposes, we consider the following resilience metric that captures a system’s capacity to absorb performance degradation, adapt to disturbances, and recover functionality within an acceptable timeframe. Let $r_k^\mathsf{des}$ denote the desired \ac{QoS} throughput for user $k$. While this target remains fixed over a coherence interval, the actual throughput at time $t$ varies due to dynamic events and is represented by $\sum_{k=1}^{K} r_k(t)$, where $r_k(t)$ is user $k$'s data rate under \ac{FBL} constraints.

Following \cite{resiliencemetric,RobertRes}, we denote three sub-metrics to quantify different aspects of system resilience. The {absorption is given by
\begin{align}\label{eq:res1}
    r_\mathsf{abs} = \frac{1}{K} \sum_{k \in \mathcal{K}} \frac{R_k(t_0)}{R_k^\mathsf{des}},
\end{align}
which captures the performance retained immediately after the disruption occurs at time \( t_0 \). The {adaptation} is defined as
\begin{align}\label{eq:res2}
    r_\mathsf{ada} = \frac{1}{K} \sum_{k \in \mathcal{K}} \frac{r_k(t_q)}{r_k^\mathsf{des}},
\end{align}
and reflects the system’s ability to restore service quality at the recovery time \( t_q \). Lastly, the time-to-recovery is expressed as
\begin{align}\label{eq:res3}
    r_\mathsf{rec} =
    \begin{cases}
    1, & \text{if } t_q - t_0 \leq T_0 \\
    \frac{T_0}{t_q - t_0}, & \text{otherwise},
    \end{cases}
\end{align}
where \( T_0 \) represents the desired maximum recovery time, beyond which the degradation is no longer considered tolerable, and thus penalized.
Subsequently, the three sum-metrics are combined into a unified resilience metric:
\begin{align}\label{eq:r}
    r = \lambda_1 r_\mathsf{abs} + \lambda_2 r_\mathsf{ada} + \lambda_3 r_\mathsf{rec},
\end{align}
where $\lambda_i \geq 0$ and $\sum_{i=1}^{3} \lambda_i = 1$, reflecting the network’s prioritization of robustness, recovery quality, and speed.
It should be noted, that compared to our previous works, the adaption metric (\ref{eq:res2}) is defined using the \ac{FBL} rate expression. This means that after the absorption, the system can still decide whether to engage adaption and switch to the \ac{FBL} regime or to ignore the disruption and continue transmitting with \ac{IBL}.

\section{Problem Formulation}
We formulate an optimization problem that incorporates the impact of \ac{FBL} on the resilience metric. To this end, we utilize the adaptation metric $r_\mathsf{ada}$ to optimize the network-wide adaptation gap $\Uppsi$, as introduced in \cite{RobertRes, RIS_RES}
\begin{align}\label{eq:Uppsi}
 \Uppsi =  \sum_{k\in \mathcal{K}} \big|\frac{r_k}{r_k^\mathsf{des}}-1\big|
\end{align}
which quantifies the system's deviation from fulfilling each user's desired quality-of-service (\ac{QoS}) target $r_k^\mathsf{des}$. Minimizing $\Uppsi$ promotes fair and reliable service adaptation across all users.

Utilizing this objective, the overall problem can be formulated as
	\begin{align}
\label{Prob1}
\tag{P1}
\underset{\vect{w},\vect{v},\vect{r}}{\min} \quad & \Uppsi \\
\text{s.t.} \quad
& \text{(\ref{eq:powConst})}, \nonumber \\
 r_k& \leq B \left(\log_2(1+\Gamma_k) - \frac{\Omega}{\sqrt{\eta}} \sqrt{V(\Gamma_k)} \right), \forall k \in \mathcal{K}, \label{rateConst} \\
 |v_m|& = 1, \quad \forall m \in \{1,\dots,M\}, \label{unit_mod}
\end{align}
where $\vect{r} = [r_1, r_2, \dots, r_K]^T$ represents the stacked rate vector, and the unit-modulus constraints in (\ref{unit_mod}) enforce the phase shift conditions $0\leq \theta_m \leq 2\pi, \forall m \in \{1,\dots,M\}$. The rate constraints in \eqref{rateConst} also include the \ac{FBL} penalty term that captures the rate degradation due to finite blocklength coding. However, as the blocklength increases, this penalty diminishes, and the achievable rate asymptotically approaches the idealized \ac{IBL} capacity. In this work, we exploit this relationship by adopting the \ac{IBL} formulation during the steady-state operation of the system prior to any disruptions. Conversely, once a disruption is detected and the system enters the adaptation phase, we optimize (\ref{Prob1}) to accurately capture the performance constraints imposed by short packet transmission during recovery.

Solving problem (\ref{Prob1}) is complex due to the interdependence between the variables $\vect{w}$ and $\vect{v}$ in the constraints in (\ref{rateConst}). Further the nature of the channel dispersion expression the unit-modulus constraint are additional challenges, especially when coupled with the rate constraints. These elements together make the optimization problem non-trivial, requiring specialized techniques such as alternating optimization and \acp{SCA} to handle these interdependencies effectively \cite{SynBenefits}. In fact, both sub-problems can be efficiently solved within the same \ac{SCA} framework \cite{accelRec}. As a result, full convergence of one sub-problem is not required before moving to the other; instead, just one iteration of each sub-problem can be performed before switching to the other. This approach significantly accelerates the overall convergence process, which is crucial in resilience scenarios and allows the evaluation of the resilience performance without the need to converge \cite{RIS_RES,accelRec}.

To facilitate the application of these techniques, we reformulate problem (\ref{Prob1}) into an equivalent but more tractable form, enabling efficient application of alternating optimization and SCA under FBL constraints as
\begin{align}\label{P1.1}
		\underset{\vect{w},\vect{r},\vect{q},\vect{u}}{\min} \quad & \Uppsi
\tag{P1.1} \\
		\text{s.t.}\quad & (\ref{eq:powConst}), (\ref{unit_mod}) \nonumber\\
		 \,\,\ r_k &\leq B \big( \log_2(1+q_k) - \frac{\Omega}{\sqrt{\eta}} u_k \big), \, &&\forall k \in \mathcal{K}, \label{P2rate}\\
		q_k  \,\,\ &\leq \Gamma_k , \, &&\forall k \in \mathcal{K}, \label{P2SINR}\\
u_k \,\,\  &\geq \sqrt{V(q_k)}  &&\forall k \in \mathcal{K},\label{gradVP1}\\
		\:\,\:\,\vect{q} \:\,&\geq 0 , \,\, \vect{u} \:\,\geq 0,\label{P2t1}
	\end{align}
where the introduction of the slack variables $\vect{q}=[q_1 ,\dots ,q_K]$ and $\vect{u}=[u_1 ,\dots ,u_K]$ convexify the rate expressions. Furthermore, (\ref{P2t1}) signifies that all values in $\vect{q}$, and also $\vect{u}$, are nonnegative. However, the constraints in (\ref{P2SINR})-(\ref{gradVP1}) remain non-convex but can be rendered in a convex form through the \ac{SCA} approach, when alternating between $\vect{w}$ and $ \vect{v}$.

\subsection{Beamforming Design}
Due to the use of alternating optimization, the phase shifts $\vect{v}$ are treated as constant during the beamforming vector design process. To obtain a convex approximation of problem (\ref{P1.1}), the constraints in (\ref{P2SINR})-(\ref{gradVP1}) need to be reformulated into a convex form.
Following \cite{RIS_RES,accelRec}, the first-order Taylor approximation around the point 	$(\tilde{\vect{w}},\tilde{\vect{q}})$ can be applied to the \ac{SINR} constraints. The resulting convex approximation of (\ref{P2SINR}) can be written as
	\begin{align}\label{SINR_convex1}
	\sum_{i\in\mathcal{K}\setminus\{k\}}|(\RisChan{k})^H \vect{w}_i|^2 + \sigma^2 +
			\frac{|(\RisChan{k})^H \tilde{\vect{w}}_k|^2}{(\tilde{q}_k)^2}q_k \nonumber \\ - \frac{2 \text{Re} \{\tilde{\vect{w}}_k^H(\RisChan{k})(\RisChan{k})^H\vect{w}_k \}}{\tilde{q_k}} \leq 0, \forall k \in \mathcal{K}.
\end{align}

Regarding the term for the channel dispersion parameter $\sqrt{V(q_k)}$ in (\ref{gradVP1}), we can derive the first-order Taylor approximation around to point $\tilde{q}_k$ \cite{Cleckx_approx} as
\begin{align}\label{eq:ChanDispApp}
 \sqrt{V(q_k)} \leq& \sqrt{1 - (1+\tilde{q}_k)^{-2}} + \big( 1+\tilde{q}_k  \big)^{-3}\nonumber\\ &(1 - (1+(\tilde{q}_k)^{-2}))^{-\frac{1}{2}} (q_k - \tilde{q}_k)  \triangleq  U_k(q_k)
\end{align}

Thus, the approximation of problem (\ref{P1.1}) can be written as
\begin{align}\label{Prob2.1}
	\underset{\vect{w},\vect{r},\vect{q},\vect{u}}{\min} \quad & \Uppsi \tag{P2} \\
	\text{s.t.}\quad & (\ref{eq:powConst}),(\ref{P2rate}), (\ref{P2t1}),  (\ref{SINR_convex1}), \nonumber\\
&u_k \geq U_k(q_k), \quad  \forall k \in \mathcal{K}. \label{chanDisp}
\end{align}
Problem (\ref{Prob2.1}) is convex and can be solved iteratively using the \ac{SCA} method. More specifically, we denote $\mat{\Lambda}_z^w = [\vect{w}_z^T ,\vect{\kappa}_z^T ]^T$ as a vector stacking the optimization variables of the beamforming design problem at iteration $z$, where $\vect{\kappa}_z = [\vect{r}_z^T,\vect{q}_z^T,\vect{u}_z]^T$. Similarly $\hat{\mat{\Lambda}}_z^w  = [\hat{\vect{w}}_z^T ,\hat{\vect{\kappa}}_z^T ]^T$ and $\tilde{\mat{\Lambda}}_z^w  = [\tilde{\vect{w}}_z^T ,\tilde{\vect{\kappa}}_z^T ]^T$ define the optimal solutions and the point, around which the approximations are computed, respectively. With these expressions, given a point $\tilde{\mat{\Lambda}}_{z}^w $, an optimal solution $\hat{\mat{\Lambda}}_z^w $ can be obtained by solving problem (\ref{Prob2.1}).

\newcommand{\RisChanv}[2]{\tilde{h}_{#1,#2} + \tilde{\mat{G}}_{#1,#2}\vect{v}}

\subsection{Phase Shift Design}
During the design of the phase shifts at the \ac{RIS}, the beamforming vectors are held fixed, following the alternating optimization strategy. To maintain consistency with the problem structure in (\ref{Prob2.1}), we define $(\RisChan{i})^H \vect{w}_k$ $=$ $\tilde{h}_{i,k}^* + \tilde{\mat{G}}_{i,k}^* \vect{v}^* = \beta_{i,k}$, where $\tilde{h}_{i,k}^* = \vect{w}_k^H \vect{h}_i$ and $\tilde{\mat{G}}_{i,k}^* = \vect{w}_k^H \mat{G}_i$, where $(\cdot)^*$
denotes the complex conjugate. With these definitions, the \ac{SINR} constraints in (\ref{P2SINR}) can be approximated using a similar procedure as applied in the beamforming design. Following the approach in \cite{RIS_RES,accelRec}, the first-order Taylor approximation around the point $(\tilde{\vect{v}}, \tilde{\vect{q}})$ can be calculated as
\begin{align}\label{apprxV}
\sum_{i\in\mathcal{K}\setminus\{k\}}&|\beta_{k,i}|^2 + \sigma^2 -\frac{|{\beta_{k,k}}|^2}{\tilde{q}_k }-\frac{2}{\tilde{q}_k} \Re\left\{ \beta^*  \vect{G} \left( \vect{v} - \tilde{\vect{v}} \right) \right\} \nonumber\\
&\qquad \qquad \quad +\frac{|\beta_{k,k}|^2}{\tilde{q}_k^2} \left( q_k - \tilde{q}_k \right) , \quad \forall k \in \mathcal{K}
\end{align}
To handle the unit-modulus constraint in (\ref{unit_mod}), we adopt the penalty method proposed in \cite{RIS_RES}. Specifically, the term $\sum_{m=1}^{M}(|v_m|^2 -1)$ is approximated using a weighted first-order Taylor expansion around a given point $\tilde{\vect{v}}$. The resulting expression,
\[
\Phi = \alpha_{v}\sum_{m=1}^{M}\text{Re}\left\{2\tilde{v}_m^*v_m - |\tilde{v}_m|^2\right\},
\]
is then incorporated into the objective function as a penalty, where $\alpha_{v} \gg 1$ is a large constant controlling the strength of the penalization.

At this point, the approximated optimization problem for the phase shift design can be formulated as
\begin{align}\label{Prob3}
	\underset{\vect{v},\vect{r},\vect{q},\vect{u}}{\min}  \,\,\, \,\,& \Uppsi - \Phi \tag{P3} \\
	\text{s.t.} \quad &(\ref{P2rate}), (\ref{P2t1}),  (\ref{SINR_convex1}),(\ref{chanDisp}),(\ref{apprxV}).
 \nonumber
\end{align}
Due to the similarity of the problem formulation and the utilization of the same \ac{SCA} framework, problem (\ref{Prob3}) can be solved by defining $\mat{\Lambda}_z^v = [\vect{v}_z^T, \vect{\kappa}_z^T]^T$ and following the same iterative procedure as for solving problem (\ref{Prob2.1}).

\subsection{Resilience-Guided Alternating Optimization}
In the context of resilience, our goal is to find a solution that meets a specific trade-off defined by the weights $\lambda_i, \forall i\in\{1,2,3\}$ in (\ref{eq:r}). At the same time, we want the system to adapt to failures quickly. Therefore, instead of fully optimizing each sub-problem until convergence, we perform just one iteration for each sub-problem before switching to the other \cite{RIS_RES,accelRec}. This is only possible because the same framework is used to solve both sub-problems, which helps us reduce the adaptation gap more rapidly. Once the adaptation gap converges, the algorithm then enhances adaptability and robustness to prepare for future failures while keeping the gap minimized. The detailed steps of the resilience-guided alternating optimization are presented in Algorithm \ref{alg}, where $T_\mathsf{calc}$ is the time needed to compute the solution of a sub-problem and $T_\mathsf{c}$ is the coherence time.
\vspace{0pt}
\begin{algorithm}[H]
\setlength{\topsep}{0pt}
\setlength{\parskip}{0pt}
\footnotesize
\caption{Resiliency-aware Alternating Optimization}\label{alg}
\begin{tikzpicture}[>={Latex[length=5pt]}, baseline=(current bounding box.north)]\label{alg:AltOpt}
	\footnotesize
	\tikzset{myRect/.style={draw,rectangle,minimum width=0.5cm, minimum height=0.25cm,align=center}}
	\tikzset{myDiam/.style={draw,diamond,aspect=1.5,minimum width=1.25cm, minimum height=0.9cm,align=center}}
	\tikzset{myArrow/.style={->,draw,line width=0.25m}}
	\tikzset{myDot/.style={draw,minimum size=0.1,circle,fill,scale=0.15}}
	
	\node[align=center](input) at (-1,-1.85){ Input:\\ $\tilde{\vect{w}},\tilde{\vect{v}},\tilde{\vect{\kappa}},$\\ $T_\mathsf{c}$,${\alpha_v}$};
	
	\node[myRect](initLam) at ($(input)+(0,2.45)$){Create\\$\tilde{\mat{\Lambda}}_z^o$};
	%\node (initLamTxt) at (initLam){};

	\node[myDiam] (o1) at (0,0){};
	\node[] (o1txt) at (o1){$o=w$};
	\node[myRect] (P2) at ($(o1)+(2.625,0.3)$){$\hat{\vect{\Lambda}}_{z+1}^w \leftarrow$ solve (P2)};
	\node[myRect] (P3) at ($(o1)+(2.625,-0.3)$){$\hat{\vect{\Lambda}}_{z+1}^v \leftarrow$ solve (P3)};
	\node[myDiam] (oT) at (6.,-0.75){};
	\node[] (oTText) at (oT){$T{<}\:T_\mathsf{c}$};
	\node[myDiam] (o2) at (4.555,-1.5){};
	\node[] (o2txt) at (o2){$o=w$};
	\node[myRect,align=center] (Lambw) at ($(o2)+(-2.45,0.45)$){$[\tilde{\vect{w}}^T, \tilde{\vect{\kappa}}^T]^T \leftarrow \hat{\mat{\Lambda}}_z^w$ ,\\$\tilde{\mat{\Lambda}}_z^v \leftarrow [\tilde{\vect{v}}^T, \tilde{\vect{\kappa}}^T]^T $};
	\node[myRect,align=center] (Lambv) at ($(o2)+(-2.45,-0.45)$){$[\tilde{\vect{v}}^T, \tilde{\vect{\kappa}}^T]^T \leftarrow \hat{\mat{\Lambda}}_z^v$ \\ $\tilde{\mat{\Lambda}}_z^w \leftarrow [\tilde{\vect{w}}^T, \tilde{\vect{\kappa}}^T]^T$};
	
%	\node[myDiam](Psi) at ($(oT)+(-0.45,-1.025)$){};
%	\node(Psitxt) at (Psi) {$\Uppsi<\tau$};

	\draw[->] (input.north) |- ($(input.north)!0.25!(initLam.240)$) -|node[pos=0.76,anchor=210]{$\,\,\,\phantom{.}_{o\leftarrow w}$}node[pos=0.535,anchor=210]{$\phantom{.}_{T{\leftarrow 0}}$}node[pos=0.655,anchor=210]{$\phantom{.}_{z{\leftarrow 0}}$} (initLam.240);
	\draw[->] (initLam.south) |- (o1.west);

	\draw[->] (o1.east) --node[myDot,pos=1]{} ($(o1.east)+(0.1,0)$) |-node[pos=0.65,above]{\tiny$\mathsf{T}$} (P2.west) ;
	\draw[->] (o1.east) -- ($(o1.east)+(0.1,0)$)  |-node[pos=0.65,above]{\tiny$\mathsf{F}$} (P3.west) ;
	
	\node[myDot] (incrementDot) at ($(oT.north)-(1.9,-0.25)$){};
	\node[myDot] (incrementDot2) at ($(oT.north)-(1.75,-0.25)$){};
    \draw     (incrementDot) -- (incrementDot2);

    \draw[->] (P2.east) -| (incrementDot);
	\draw[->] (P3.east) -| (incrementDot);
	
	\draw[->] (incrementDot2) |-node[pos=0.775,below]{$\phantom{.}_{z\,{\leftarrow}{z+1}}$}node[pos=0.775,above]{ \tiny${T{\leftarrow}{T\hspace{-0.075cm}+\hspace{-0.075cm}T_\mathsf{calc}}}$} (oT.west);
\draw[->] (incrementDot2) |-node[pos=1,anchor=180]{Output:\\$\hat{\vect{w}}_z,\hat{\vect{v}}_z$}($(incrementDot2)+(0.5,0.5)$) ;
	
	%\draw[->] (oT.north) |- node[pos=0.1,right]{\tiny$\mathsf{T}$} node[pos=0.65,below]{$\tilde{\mat{\Lambda}}_z^o \leftarrow \hat{\mat{\Lambda}}_z^o$}  ($($(oT.north)!0.5!(o1.north)$) + (0,0.4) $) -| (o1.north);
	
	\draw[->] (oT.south) |- node[pos=0.35,anchor=-35]{\tiny$\mathsf{F}$}  (o2.east); %($(oT.west)!0.51!(o2.north)$)
	
	\draw[->] (o2.west) --node[myDot,pos=1]{} ($(o2.west)-(0.125,0)$) |-node[pos=0.65,above]{\tiny$\mathsf{T}$} (Lambw.east) ;
	\draw[->] (o2.west) -- ($(o2.west)-(0.125,0)$)  |-node[pos=0.65,above]{\tiny$\mathsf{F}$} (Lambv.east) ;
	
	\node[myDot] (reassignDot) at  ($(o2)-(4.39,0)$) {};
	\draw[->] (Lambw.west) -|node[pos=0.25,above]{$\phantom{.}_{o\leftarrow v}$} (reassignDot) ;
    \draw[->] (Lambv.west) -|node[pos=0.25,above]{$\phantom{.}_{o\leftarrow w}$} (reassignDot) ;

    \draw[->] (reassignDot) -|node[pos=0.65,anchor=-25]{} (o1.south);
	
	\path (oT.east) |-node[pos=0.85,anchor=125,align=center](a){} node[pos=0.15,anchor=35,align=center](b){\tiny$\mathsf{T}$} ($(oT.east)-(0.0,1)$);% -- (Psi.east); %(oT.east)!0.5!
	
\draw[->] (oT.east) -- ($(oT.east)-(0.0,1)$);
	\node[align=center] at ($(a)+(-0.1,-0.15)$){Stop};
	
%Output:\\$\hat{\vect{w}}_z,\hat{\vect{v}}_z$
%	\draw[->] (Psi.south) --node[pos=0.25,right]{\tiny$\mathsf{T}$} ($(Psi.south)+(0,-0.4)$);
	
%	\node at ($(Psi.south)+(0,-0.65)$) {Stop};
	
\end{tikzpicture}
\end{algorithm}

\section{Numerical results}
In this section, the performance of our proposed finite blocklength algorithm within the  resilience framework is evaluated. To this end we assume a cell-free \ac{MIMO} system with $N=3$ \acp{AP}, each of which equipped with $L=8$ antennas and positioned in the center of  random quadrants of the area, which spans $[-500,500] \times [-500,500] \text{m}^2$. We assume $K=6$ single-antenna users to be randomly distributed within the quadrants that are equipped with an \ac{AP}. In the center of the served area a \ac{RIS} is deployed, consisting of \( M = 1000 \) reflective elements (if not specified otherwise) arranged in a quadratic grid with \( \lambda_f/4 \) spacing, where \( \lambda_f = 0.1 \)~m is the wavelength. For the RIS channels, we use the correlated channel model from \cite{corrBj}. Further, reflected channels follow a line-of-sight model, while direct channels experience Rayleigh fading with log-normal shadowing (8~dB standard deviation). Moreover, we assume a bandwidth of $B=10$ ~{MHz}, a noise power of $\sigma^2 = -100$~dBm, a maximum transmit power of $P_n^\mathsf{max} = 32$~dBm, $T_0=5$~s and each user to request a \ac{QoS} of $r_k^\mathsf{des}=37$~Mbps.
The system is affected by an outage, which is defined as the event where any direct link between \ac{AP} $n$ and user $k$ can be subjected to a full blockage, effectively removing the link from the network. Because the \ac{RIS} is positioned to bypass potential blockages the RIS-assisted links remain unobstructed. Without loss of generality, we assume that the blockage affects the strongest direct channel link in order to force a reallocation of the system's resources. During such a disruption, the system faces a critical decision: whether to initiate recovery procedures, thereby attempting to compensate both for the degradation due to the blockage and the rate penalty introduced by \ac{FBL} transmission, or to ignore the outage and continue operating under the degraded conditions. To highlight these effects, we prioritize adaptation and time-to-recovery by choosing the resilience weights as \( \lambda_1 = 0.1 \), \( \lambda_2 = 0.5 \), and \( \lambda_3 = 0.4 \).

Figure~\ref{compareBL} compares these two strategies across varying \ac{QoS} requirements $r^\mathsf{des}_k$. As shown, for $r^\mathsf{des}_k = 35$~Mbps, the system can overcome the FBL-induced rate loss and restore service performance once the blocklength exceeds approximately 60. However, when the desired rate increases to $r^\mathsf{des}_k = 37$~Mbps, the system becomes more resource-constrained and is only able to recover effectively when the blocklength reaches at least 150 symbols. Notably, for $r^\mathsf{des}_k = 40$~Mbps, the system fails to reconfigure itself sufficiently to fully compensate the FBL penalty, even when the blocklength is increased to 3000. This behavior underscores a fundamental resilience decision, in which it may be more sensible to ignore the outage and continue serving the remaining users, rather than attempting a costly recovery that could degrade overall system performance.

For the cases with \( r_k^\mathsf{des} = 35 \) and \( 37 \) Mbps, Figure~\ref{compareBL} also reveals the existence of critical turning points with respect to the blocklength \(\eta\). At these thresholds, the resilience metric \( r \) exhibits a steep increase, which represents a sudden transition from an completely unrecovered to a partially recovered rate for the blocked user. At these thresholds, the system is capable of simultaneously compensating for the finite blocklength penalty as well as recovering from the disruption, without degrading the performance of other users. Specifically, the system reaches this balance at a lower blocklength of 838 symbols when \( r_k^\mathsf{des} = 35 \) Mbps, compared to 907 symbols when \( r_k^\mathsf{des} = 37 \) Mbps. This indicates that the additional headroom in resource availability at lower \ac{QoS} demands enables the system to recover using shorter blocklengths. This becomes an important advantage in the context of resilience, where rapid reconfiguration of the system's resources is required.

To further explore the system's resilience behavior under finite blocklength constraints, we now shift focus to Figure~\ref{R_compareBL}, which maintains the same target rate of \( r_k^\mathsf{des} = 37 \) Mbps (corresponding to the red dashed curve in both figures), but varies the number of \ac{RIS} elements with \( M = 625, 1000, \) and \( 1600 \). A qualitatively similar trend is observed in this figure, where increasing the size of the \ac{RIS} configuration expedites the recovery from the \ac{FBL}-induced performance loss. For instance, both \( M = 1000 \) and \( M = 1600 \) enable the system to overcome the \ac{FBL} penalty and exhibit the same sharp resilience transition where it can begin to recover from the disruption. In contrast, when \( M = 625 \), the system fails to compensate for the \ac{FBL} rate loss entirely, remaining in a degraded state. Notably, increasing the number of reflecting elements not only improves the system’s robustness, by providing alternating paths, but also enables recovery at much shorter blocklengths. This indicates that larger RIS arrays enhance both spatial redundancy and temporal responsiveness, making them particularly effective for rapid reconfiguration in resilience-critical scenarios.

\looseness-1
\begin{figure}
  \centering
  \includegraphics[width=1\linewidth]{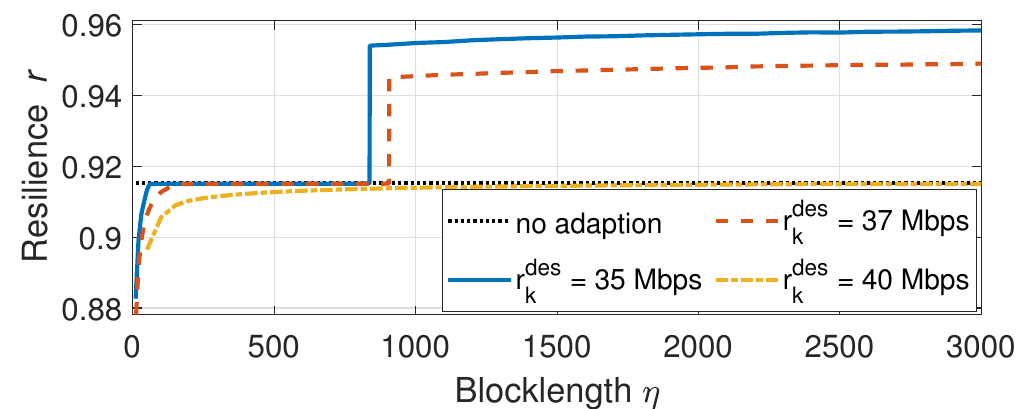}
  \caption{Resilience $r$ over the blocklength $\eta$ for different $r^\mathrm{des}_k$ with $M=1000$ RIS elements. \vspace{-0.35cm}}
  \label{compareBL}
\end{figure}

\begin{figure}
  \centering
  \includegraphics[width=1\linewidth]{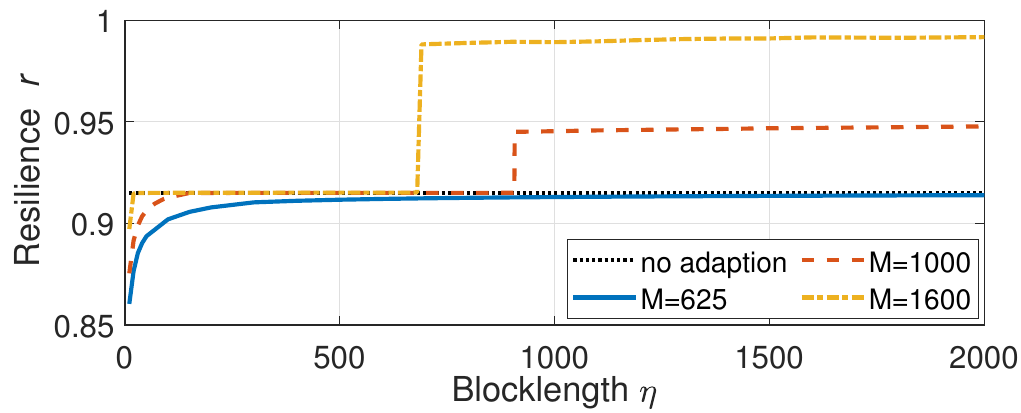}
  \caption{Resilience $r$ over the blocklength $\eta$ for different number of RIS elements with $r^\mathrm{des}_k = 37$ Mbps. \vspace{-0.35cm}}
  \label{R_compareBL}
\end{figure}
%\begin{figure}
%  \centering
%  \includegraphics[width=0.75\linewidth]{figures/ResilienceOverr.eps}
%  \caption{Overall resilience $r$ for the second blockage over the number of reflecting elements $M$ for the proposed, baseline and robustness-only approach\vspace{-1.5cm}}
%  \label{overallR}
%\end{figure}
\looseness-1

\section{Conclusion}\label{ch:conc}
As wireless networks become foundational for critical operations, ensuring their resilience by rapidly adapting to unforeseen disruptions is essential. In this work, we proposed a resilience framework that incorporates finite blocklength-aware recovery and leverages RIS to enhance adaptation potential and mitigate finite blocklength penalties. Our approach balances absorption, adaptation, and time-to-recovery, enabling a clear comparison between system performance when reallocating resources versus operating under degraded conditions. Numerical results demonstrated critical blocklength thresholds that distinguish between compensating solely for finite blocklength losses and fully recovering from both finite blocklength penalties and the initial disruption. Furthermore, larger RIS deployments were shown to accelerate recovery and enable operation with shorter blocklengths, emphasizing the trade-offs between rate, blocklength, and reconfiguration effort. These findings highlight the importance of resource surplus and reconfigurable intelligent surface size in supporting rapid and effective recovery under stringent latency and \ac{QoS} demands. In fact, even a one-symbol increase in blocklength can determine the success or failure of recovery, highlighting the sensitivity of resilient operation. Overall, our framework provides valuable insights and a practical foundation for designing wireless systems that sustain uninterrupted performance in dynamic and complex environments.

\balance
%\newpage
%\pagestyle{scrplain}
%\appendix
%
%\input{content/appendix}

%\cleardoublepage

%\cleardoublepage

%%

%\addcontentsline{toc}{chapter}{List of Figures}
%\listoffigures
%
%\listofalgorithms
%\cleardoublepage

%\addcontentsline{toc}{chapter}{List of Tables}
%\listoftables
%\cleardoublepage

%%
%\cleardoublepage

%\renewcommand*{\lstlistlistingname}{List of Listings}
%\lstlistoflistings
%\cleardoublepage

%\addcontentsline{toc}{chapter}{List of Formulas}
%\listofmyequations
%\cleardoublepage

%%

%\flushbottom
%\nocite*{}
% modified: /usr/share/texmf-texlive/bibtex/bst/dinat/dinat.bst

\footnotesize
\bibliographystyle{IEEEtran}
\bibliography{references}
\balance
%\section{Acronyms}
\begin{acronym}
\setlength{\itemsep}{0.1em}
\acro{6G}{sixth-generation}
\acro{AF}{amplify-and-forward}
\acro{AI}{artificial intelligence}
\acro{AP}{access point}
\acro{AWGN}{additive white Gaussian noise}
\acro{B5G}{Beyond 5G}
\acro{BLER}{Block Error Rate}
\acro{BS}{base station}
\acro{CB}{coherence block}
\acro{CE}{channel estimation}
\acro{C-RAN}{Cloud Radio Access Network}
\acro{CMD}{common message decoding}
\acro{CP}{central processor}
\acro{CSI}{channel state information}
\acro{CRLB}{Cramér-Rao lower bound}
\acro{D2D}{device-to-device}
\acro{DC}{difference-of-convex}
\acro{DFT}{discrete Fourier transformation}
\acro{DL}{downlink}
\acro{FBL}{finite blocklength}
\acro{GDoF}{generalized degrees of freedom}
\acro{IBL}{infinite blocklength}
\acro{IC}{interference channel}
\acro{i.i.d.}{independent and identically distributed}
\acro{IRS}{intelligent reflecting surface}
\acro{IoT}{Internet of Things}
\acro{LoS}{line-of-sight}
\acro{LSF}{large scale fading}
\acro{KPI}{key performance indicator}
\acro{M2M}{Machine to Machine}
\acro{MISO}{multiple-input and single-output}
\acro{MIMO}{multiple-input and multiple-output}
\acro{MRT}{maximum ratio transmission}
\acro{MRC}{maximum ratio combining}
\acro{MSE}{mean square error}
\acro{NOMA}{non-orthogonal multiple access}
\acro{NLoS}{non-line-of-sight}
\acro{PSD}{positive semidefinite}
\acro{QCQP}{quadratically constrained quadratic programming}
\acro{QoS}{quality-of-service}
\acro{RF}{radio frequency}
\acro{RC}{reflect coefficient}
\acro{RIS}{reconfigurable intelligent surface}
\acro{RS-CMD}{rate splitting and common message decoding}
\acro{RSMA}{rate-splitting multiple access}
\acro{RS}{rate splitting}
\acro{SCA}{successive convex approximation}
\acro{SDP}{semidefinite programming}
\acro{SDR}{semidefinite relaxation}
\acro{SIC}{successive interference cancellation}
\acro{SINR}{signal-to-interference-plus-noise ratio}
\acro{SOCP}{second-order cone program}
\acro{SVD}{singular value decomposition }
\acro{TIN}{treating interference as noise}
\acro{TDD}{time-division duplexing}
\acro{TSM}{topological signal management}
\acro{UHDV}{Ultra High Definition Video}
\acro{UL}{uplink}
\acro{w.r.t.}{with respect to}

\acro{AF}{amplify-and-forward}
\acro{AWGN}{additive white Gaussian noise}
\acro{B5G}{Beyond 5G}
\acro{BS}{base station}
\acro{C-RAN}{Cloud Radio Access Network}
\acro{CSI}{channel state information}
\acro{CMD}{common-message-decoding}
\acro{CM}{common-message}
\acro{CoMP}{coordinated multi-point}
\acro{CP}{central processor}
\acro{D2D}{device-to-device}
\acro{DC}{difference-of-convex}
\acro{EE}{energy efficiency}
\acro{IC}{interference channel}
\acro{i.i.d.}{independent and identically distributed}
\acro{IRS}{intelligent reflecting surface}
\acro{IoT}{Internet of Things}
\acro{LoS}{line-of-sight}
\acro{LoSC}{level of supportive connectivity}
\acro{M2M}{Machine to Machine}
\acro{NOMA}{non-orthogonal multiple access}
\acro{MISO}{multiple-input and single-output}
\acro{MIMO}{multiple-input and multiple-output}
\acro{MMSE}{minimum mean squared error}
\acro{MRT}{maximum ratio transmission}
\acro{MRC}{maximum ratio combining}
\acro{NLoS}{non-line-of-sight}
\acro{PA}{power amplifier}
\acro{PSD}{positive semidefinite}
\acro{QCQP}{quadratically constrained quadratic programming}
\acro{QoS}{quality-of-service}
\acro{RF}{radio frequency}
\acro{RRU}{remote radio unit}
\acro{RS-CMD}{rate splitting and common message decoding}
\acro{RS}{rate splitting}
\acro{SDP}{semidefinite programming}
\acro{SDR}{semidefinite relaxation}
\acro{SIC}{successive interference cancellation}
\acro{SCA}{successive convex approximation}
\acro{SINR}{signal-to-interference-plus-noise ratio}
\acro{SOCP}{second-order cone program}
\acro{SVD}{singular value decomposition }
\acro{TP}{transition point}
\acro{TIN}{treating interference as noise}
\acro{UHDV}{Ultra High Definition Video}
\acro{LoSC}{level of supportive connectivity}
%\acro{M2M}{Machine to Machine}
%\acro{B5G}{Beyond 5G}
%\acro{CP}{Central Processor}
%\acro{IRS}{Intelligent Reflecting Surface}
%\acro{IoT}{Internet of Things}
%\acro{BS}{base station}
%\acro{C-RAN}{Cloud Radio Access Network}
%\acro{TIN}{treating interference as noise}
%\acro{RS}{rate splitting}
%\acro{CMD}{common message decoding}
%\acro{RS-CMD}{rate splitting and common message decoding}
%\acro{UHDV}{Ultra High Definition Video}
%\acro{LoS}{line-of-sight}
%\acro{NLoS}{non-line-of-sight}
%\acro{AF}{amplify-and-forward}
%\acro{RF}{radio frequency}
%\acro{QoS}{quality-of-service}
%\acro{QCQP}{quadratically constrained quadratic programming}
%\acro{DC}{difference-of-convex}
%\acro{IC}{interference channel}
%\acro{SIC}{Successive Interference Cancellation}
%\acro{AWGN}{Additive White Gaussian Noise}
%\acro{SINR}{signal-to-interference-plus-noise
%ratio}
%\acro{SINRs}{signal-to-interference-plus-noise
%ratios}
%\acro{MRC}{maximum ratio combining}
%\acro{D2D}{device-to-device}
%\acro{MIMO}{multiple-input and multiple-output}
%\acro{i.i.d.}{independent and identically distributed}
%\acro{SOCP}{second-order cone program}
%\acro{SDR}{semidefinite relaxation}
%\acro{SDP}{semidefinite programming}
%\acro{PSD}{positive semidefinite}
%\acro{SVD}{singular value decomposition}
\end{acronym}

\balance
\end{document}